\documentclass[iop]{emulateapj}
\usepackage{amsmath}
\usepackage{natbib}
\usepackage{color}

\renewcommand{\vec}[1]{{\bf #1}}
\newcommand{\matrx}[1]{
    \ifmmode{\underline{\underline {\bf #1}}}
    \else ${\underline{\underline {\bf #1}}}$
    \fi}
\newcommand{\totd}[1]{
    \ifmmode{ {{\partial{#1}}\over {\partial t}} + \vec{v}\cdot\nabla #1}
    \else${ {{\partial{#1}}\over {\partial t}} + \vec{v}\cdot\nabla #1}$
    \fi}

\def\andrii#1{{\color[rgb]{1.0,0.0,1.0}{#1}}}

\newcommand{\figproblem}{
\begin{figure*}[ht]
  \centering
  \includegraphics[width=0.8\linewidth]{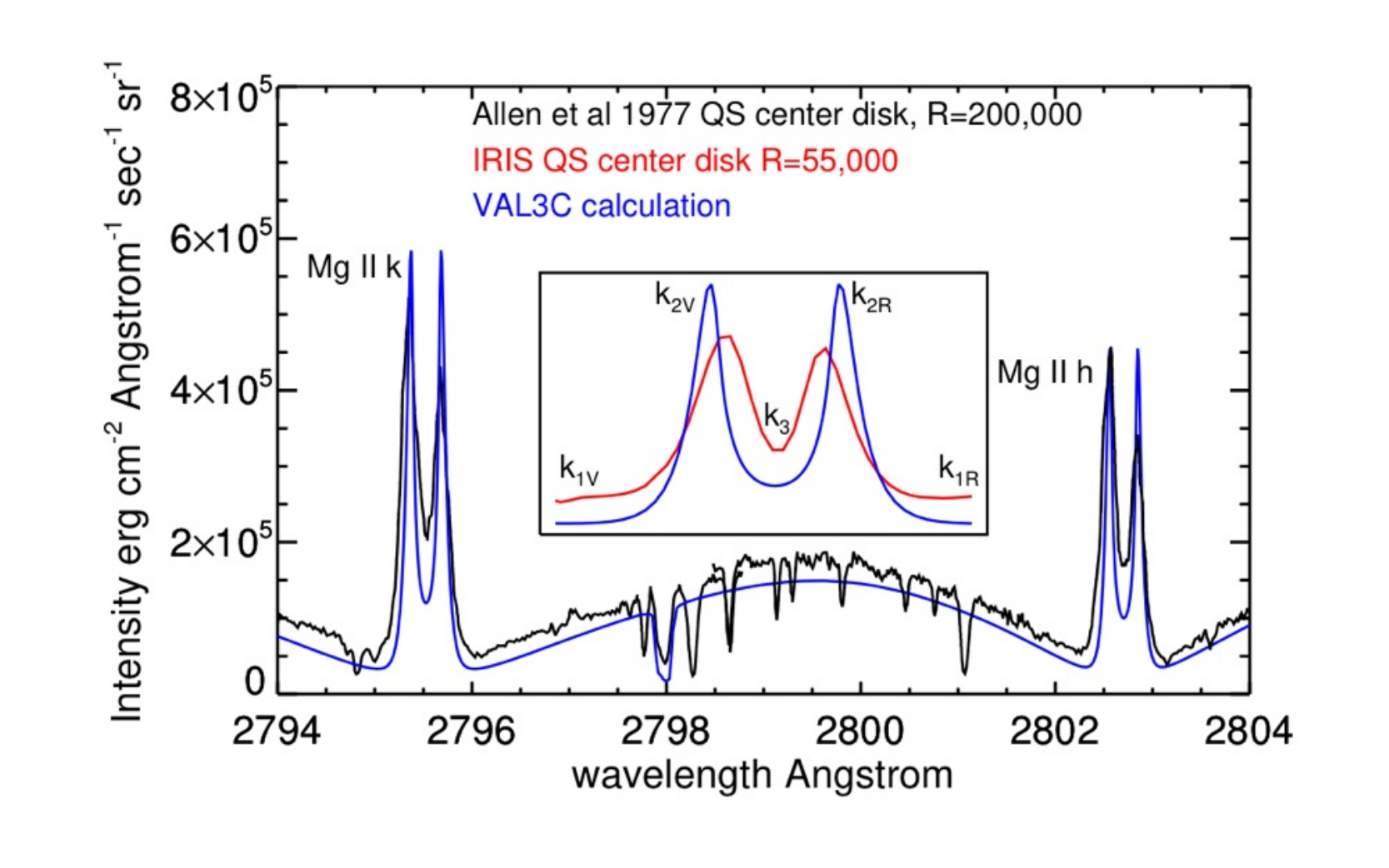}
  \caption{The figure shows observed and computed profiles of the \ion{Mg}{2}
    $h$ and $k$ lines for quiet Sun conditions close to disk center.
    Calculations were made with a full treatment of partial redistribution using
    the code RH
    \citep{Uitenbroek2001}.
    The inset shows the $k$ line with notations for the $k_1$ minima, $k_2$
    maxima, and $k_3$ minimum in intensity, discussed in the text.
    The present paper focuses on the systematic discrepancies between
    observation and theory at wavelengths \textit{between} the maxima, such as
    the $k_2$ peaks shown for \ion{Mg}{2}.
    These generally lie less than 0.30~\AA{} from line centers, of profiles of
    the resonance lines of \ion{H}{1}, \ion{Mg}{2}, and \ion{Ca}{2}, the
    strongest lines in the solar spectrum.}
  \label{fig:problem}
\end{figure*}
}

\newcommand{\figval}{
\begin{figure*}[ht]
  \centering
  \includegraphics[width=\linewidth]{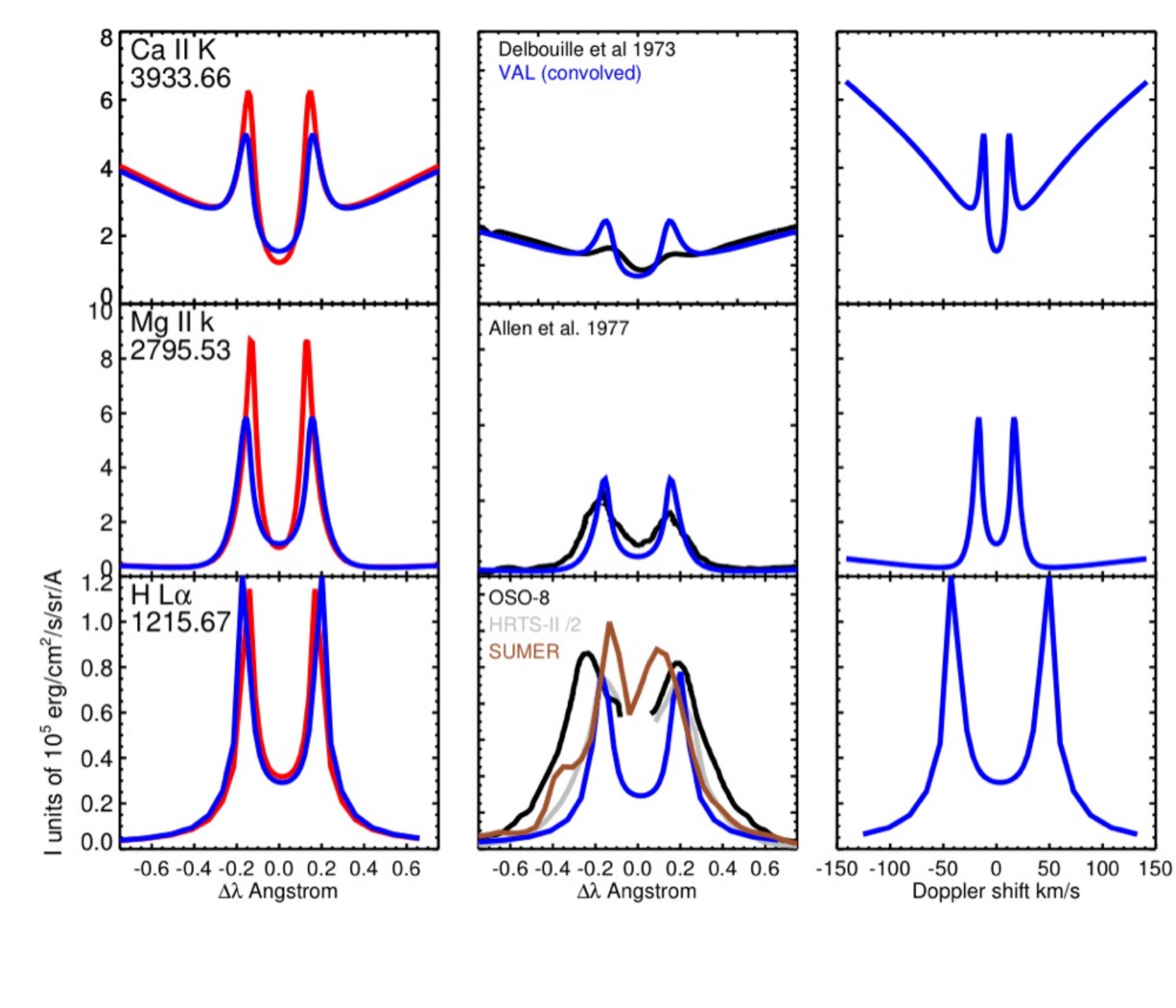}
  \caption{The figure shows the resonance lines of \ion{Ca}{2}, \ion{Mg}{2} and
    \ion{H}{1}, for two sets of calculations in model 3C of 
    \citet{Vernazza+Avrett+Loeser1981}.
    The blue lines use the micro-turbulence values taken from the
    model.
    Red lines  show calculations using a micro-turbulence with a maximum of
    1~\velu{}.
    Calculations were made with a full treatment of partial redistribution using
    the code RH
    \citep{Uitenbroek2001}.
    The middle panels compare the calculations convolved with instrumental 
     profiles. For the L$\alpha$ line the computed profiles
    were convolved with the OSO-8 profile.
    The rightmost panels show computations for VAL3-C plotted on a Doppler
    scale.}
  \label{fig:val3c}
\end{figure*}
}

\newcommand{\cartoon}{
\begin{figure}[ht]
  \centering
  \includegraphics[width=\linewidth]{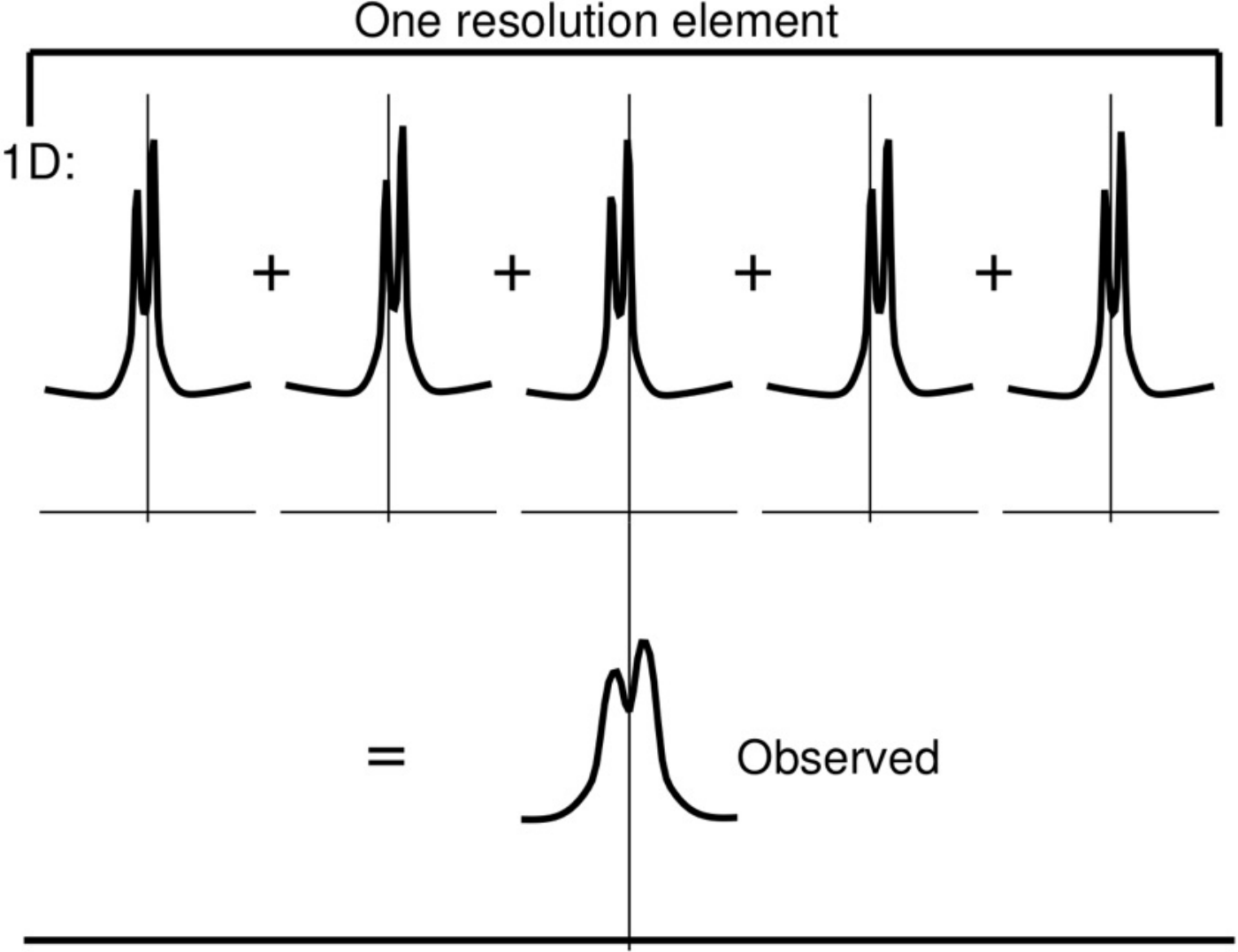}
  \caption{A schematic picture of the ``macroturbulent'' explanation for the
    reduced peak-core intensity contrast in chromospheric resonance lines.
    Each profile in the upper panel comes from an unresolved individual
    atmosphere computed in 1D, having a large peak-core ratio
    (Figure~\ref{fig:val3c}).
    In the macro-turbulent picture, these profiles presumed to be shifted in
    wavelength by an amount comparable to the core-peak wavelength difference.
    The sum of these unresolved profiles yields the profile labeled
    ``Observed''.}
  \label{fig:cartoon}
\end{figure}
}

\newcommand{\figsji}{
\begin{figure}[ht]
  \centering
  \includegraphics[width=\linewidth]{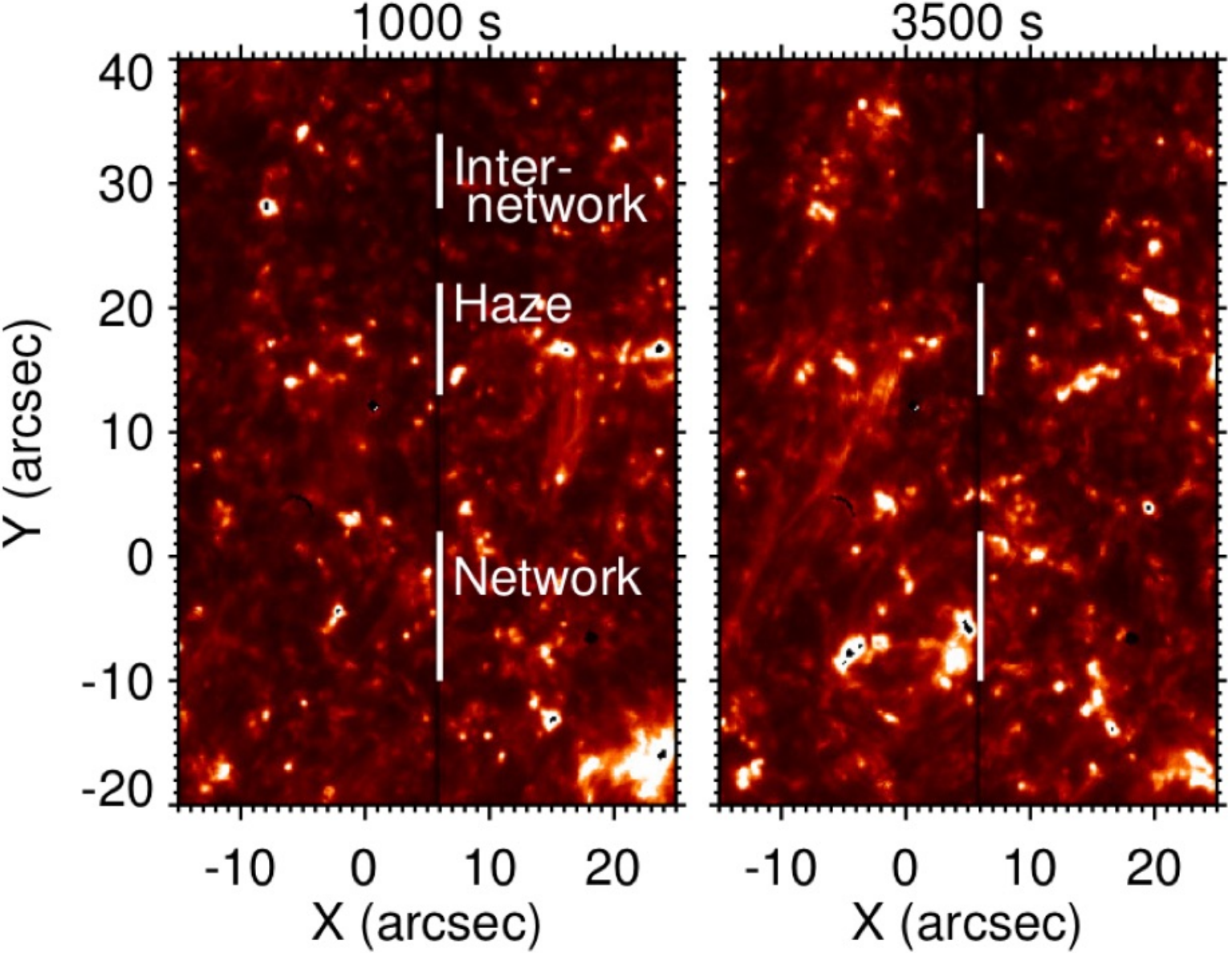}
  \caption{Slit-jaw images from the 1400~\AA{} channel are plotted to
    over-expose the brightest pixels, revealing bright regions associated with
    continuum UV emission formed in the chromospheric network.
    The two frames correspond to the times shown on the $y-$axis of the slit spectra
    (Figure~\ref{fig:irisyt}).}
  \label{fig:sji}
\end{figure}
}

\newcommand{\figyt}{
\begin{figure}[ht]
  \centering
  \includegraphics[width=\linewidth]{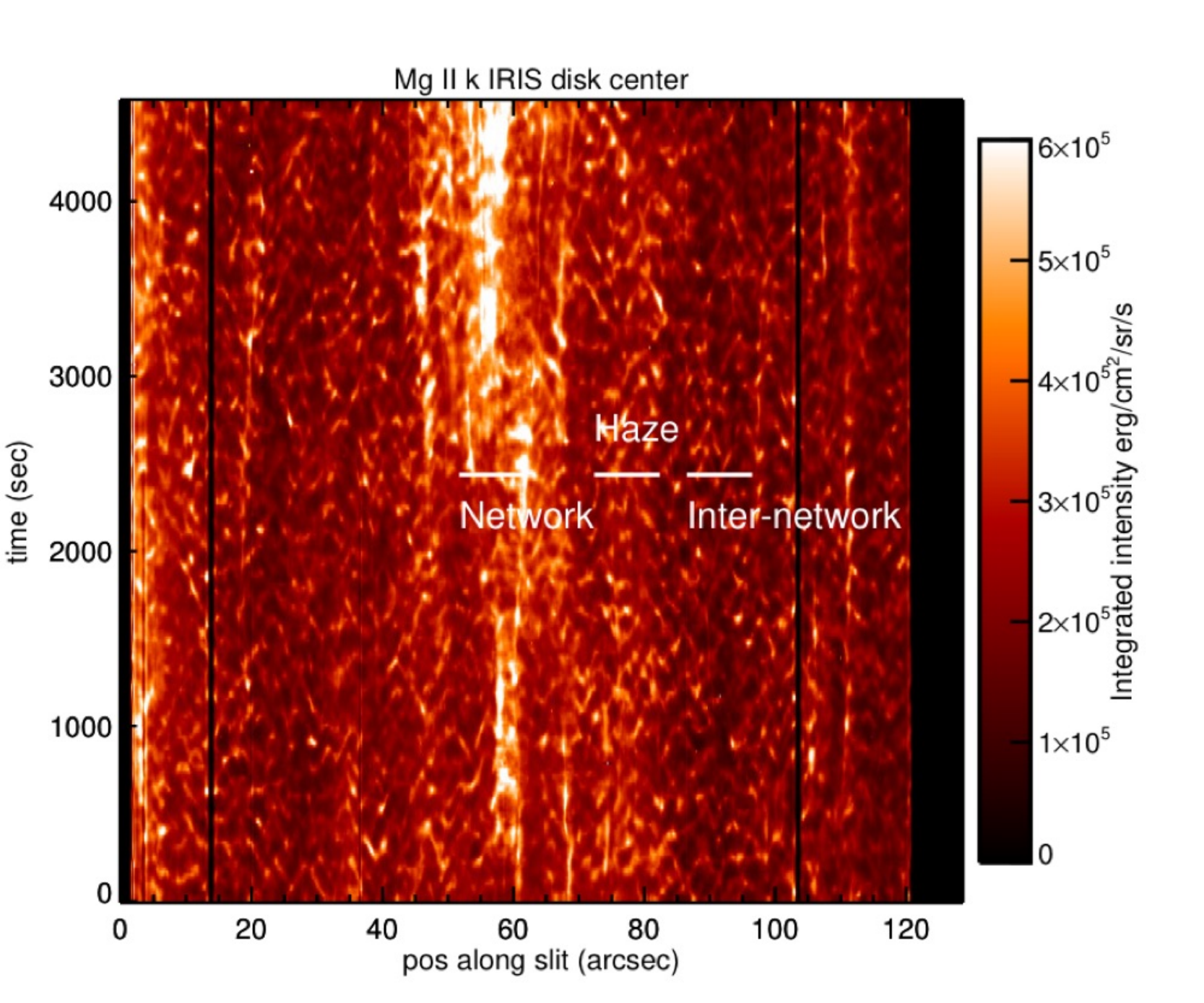}
  \caption{The wavelength-integrated intensity of the core of the \ion{Mg}{2}
    $k$-line is shown as a function of position along the IRIS slit and time,
    highlighting areas of network and inter-network.
    The area marked as ``haze'' is an intermediate region, perhaps with some
    intensity contributions from high-lying chromospheric fibrils.}
  \label{fig:irisyt}
\end{figure}
}

\newcommand{\figirismacrotk}{
\begin{figure*}[th]
  \centering
  \includegraphics[width=0.8\linewidth]{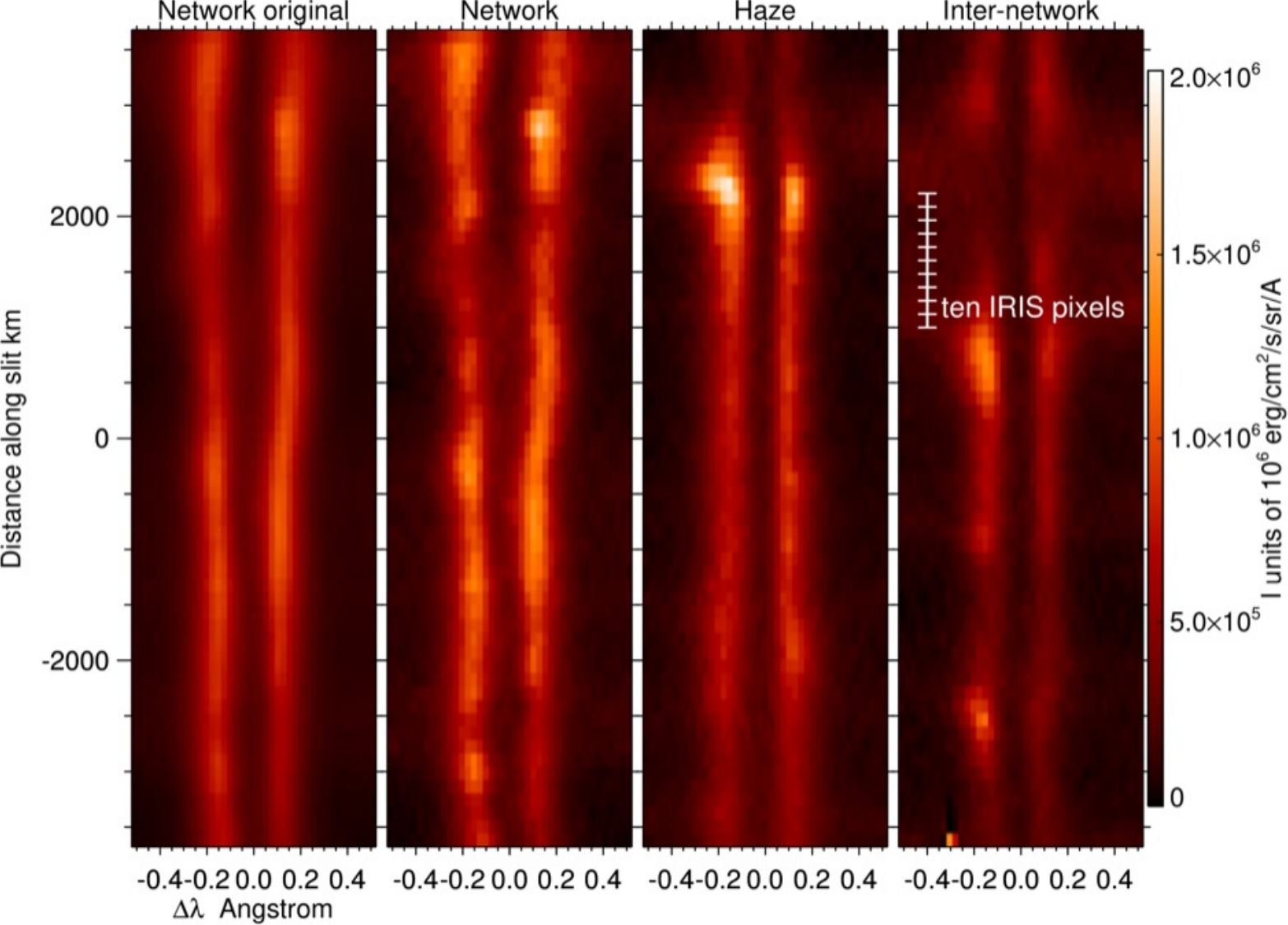}
  \caption{Line core profiles of \ion{Mg}{2} $k$ are shown as functions of
    wavelength from line center and position along the IRIS slit for the middle
    exposure of the time series.
    The left-most panel shows original data, the three right panels have been
    spatially de-convolved using the prescription of
    \citet{2018SoPh..293..125C}.
    The profiles are taken from the slit positions and time corresponding to the
    white lines shown in Fig.~\ref{fig:irisyt}.}
  \label{fig:irismacrotk}
\end{figure*}
}

\newcommand{\figiris}{
\begin{figure*}[ht]
  \centering
  \includegraphics[width=0.8\linewidth]{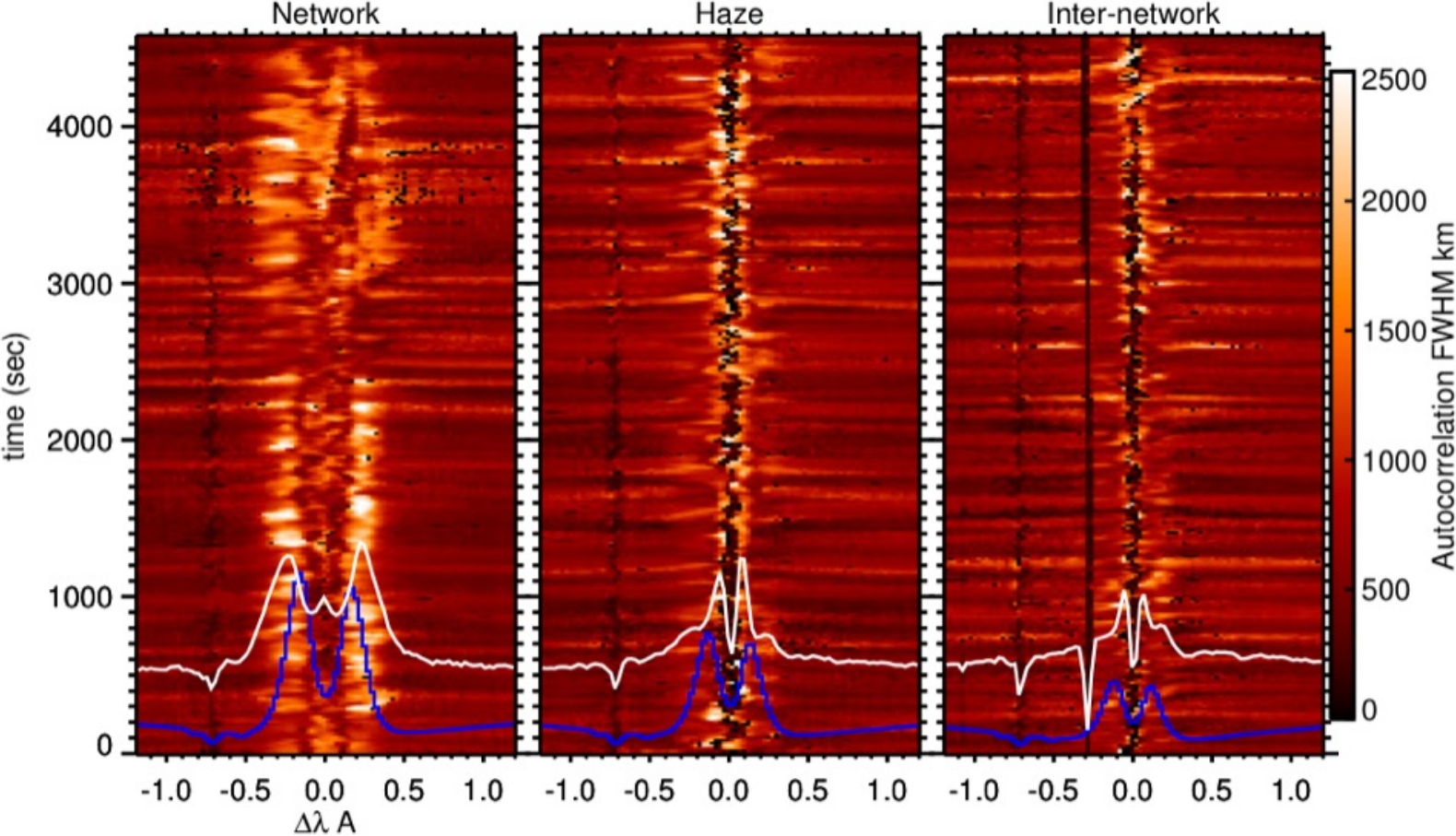}
  \caption{The spatial auto-correlation full-widths of the three regions
    (Figure~\ref{fig:irisyt}) are shown as \new{images as a function of wavelength and
    time, each ``image pixel''
    has the correlation length shown on the color bar to the right.  The  
correlation lengths shown were     averaged over the three regions:    network, haze and inter-network in Fig.~\ref{fig:irisyt}.
    The white line plots autocorrelation FWHMs also  }averaged in time along each column, the
    time scale in seconds on the left applies to the FWHM widths measured in km.
    The blue lines show, for reference, the average line profiles along the same
    columns, divided by $10^3$ (i.e. multiply the $y$-values in blue by $10^3$
    to get physical intensities in erg~cm$^{-2}$s$^{-1}$sr$^{-1}$\AA$^{-1}$).
    There is a dark pixel at $-0.3$~\AA{} in the internetwork region, the other
    features are solar.}
  \label{fig:iris}
\end{figure*}
}

\newcommand{\figmgii}{
\begin{figure*}[ht]
  \centering
  \includegraphics[width=\linewidth]{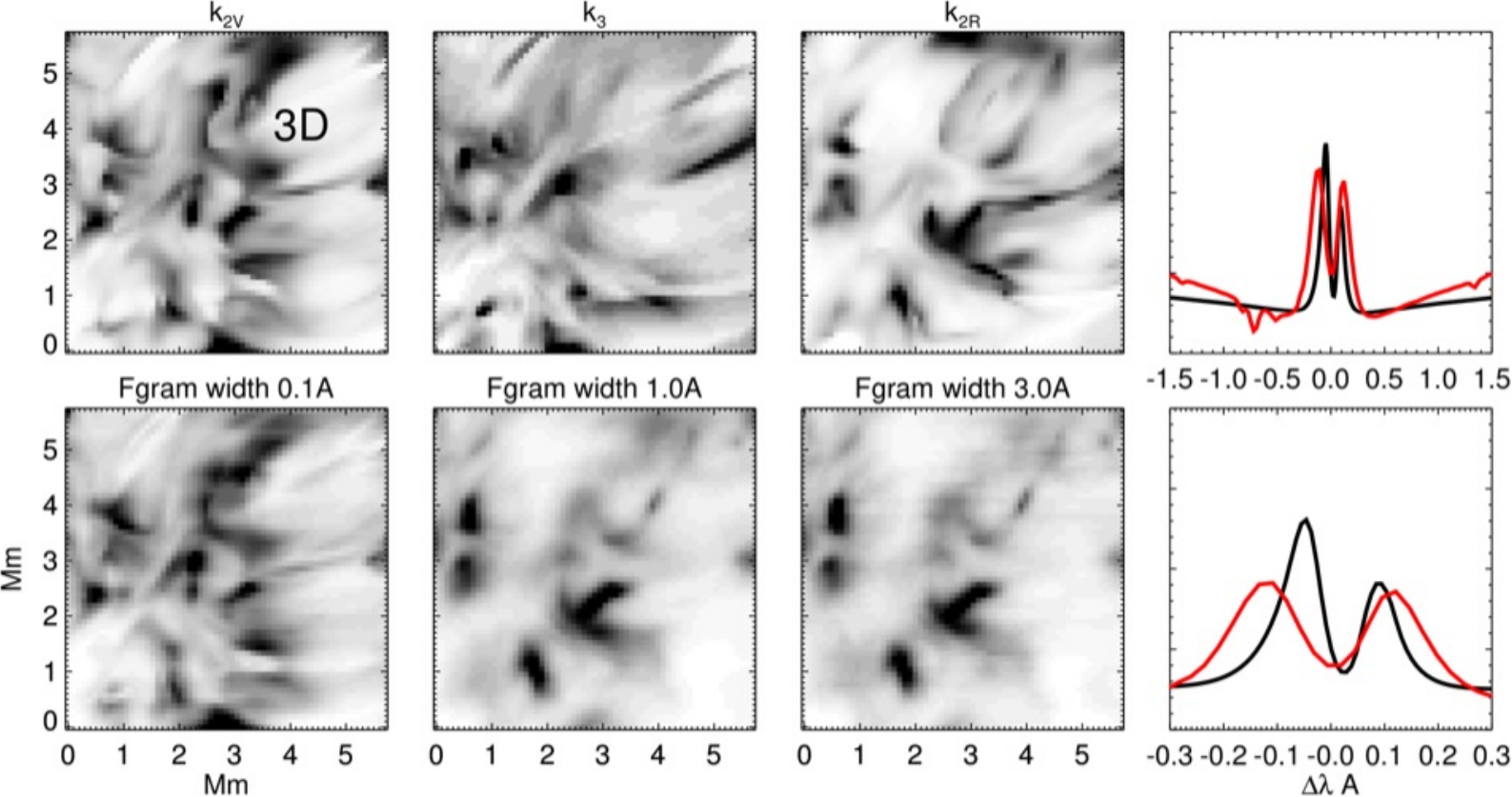}\\
  \includegraphics[width=\linewidth]{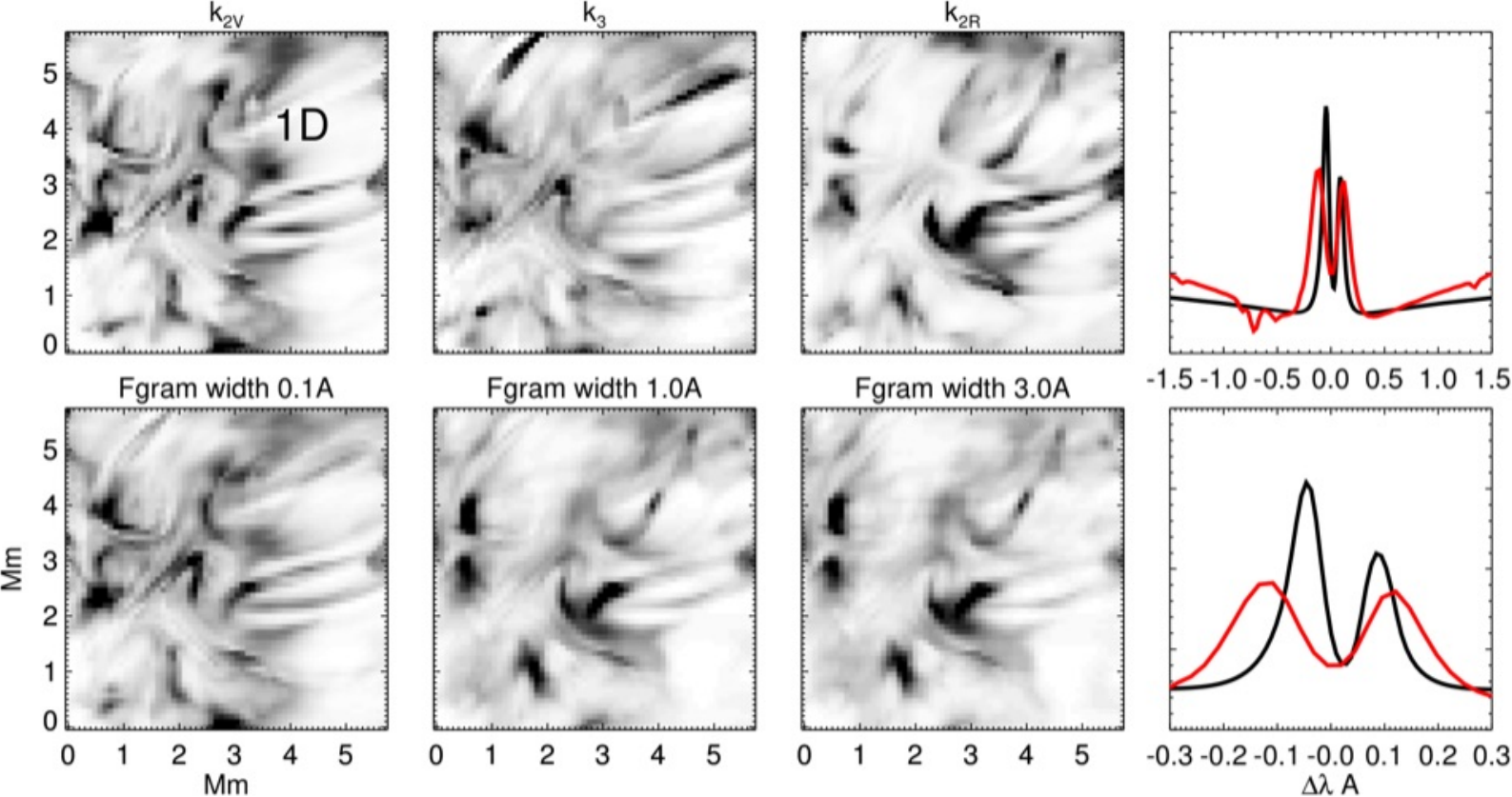}
  \caption{Images in the \ion{Mg}{2}~$k$ line computed by
    \citet{2017A&A...597A..46S}
    are shown at the computed $k_{2V}$, $k_3$ and $k_{2R}$ wavelengths (rows one
    and three), and through a filtergram instrument with the widths indicated.
    The gray scale is linear from zero to \new{ $3.8\cdot 10^{6}$
    erg~cm$^{-2}$s$^{-1}$sr$^{-1}\AA^{-1}$,} reversed (black is the maximum
    intensity) to reveal fainter emission between bright regions.
    The rightmost panels show the average spectra (black) superposed with the
    average quiet Sun from the IRIS data studied here in red.
    Although the R-MHD calculation is for a region of network, the
    region is dominated in area by non-network emission, thus we compare with
    the quiet IRIS data.
    Note that the computed line $k_2$ peak separations are
    substantially smaller than observed.}
  \label{fig:mgii}
\end{figure*}
}

\newcommand{\figautoc}{
\begin{figure}[ht]
  \centering
  \includegraphics[width=\linewidth]{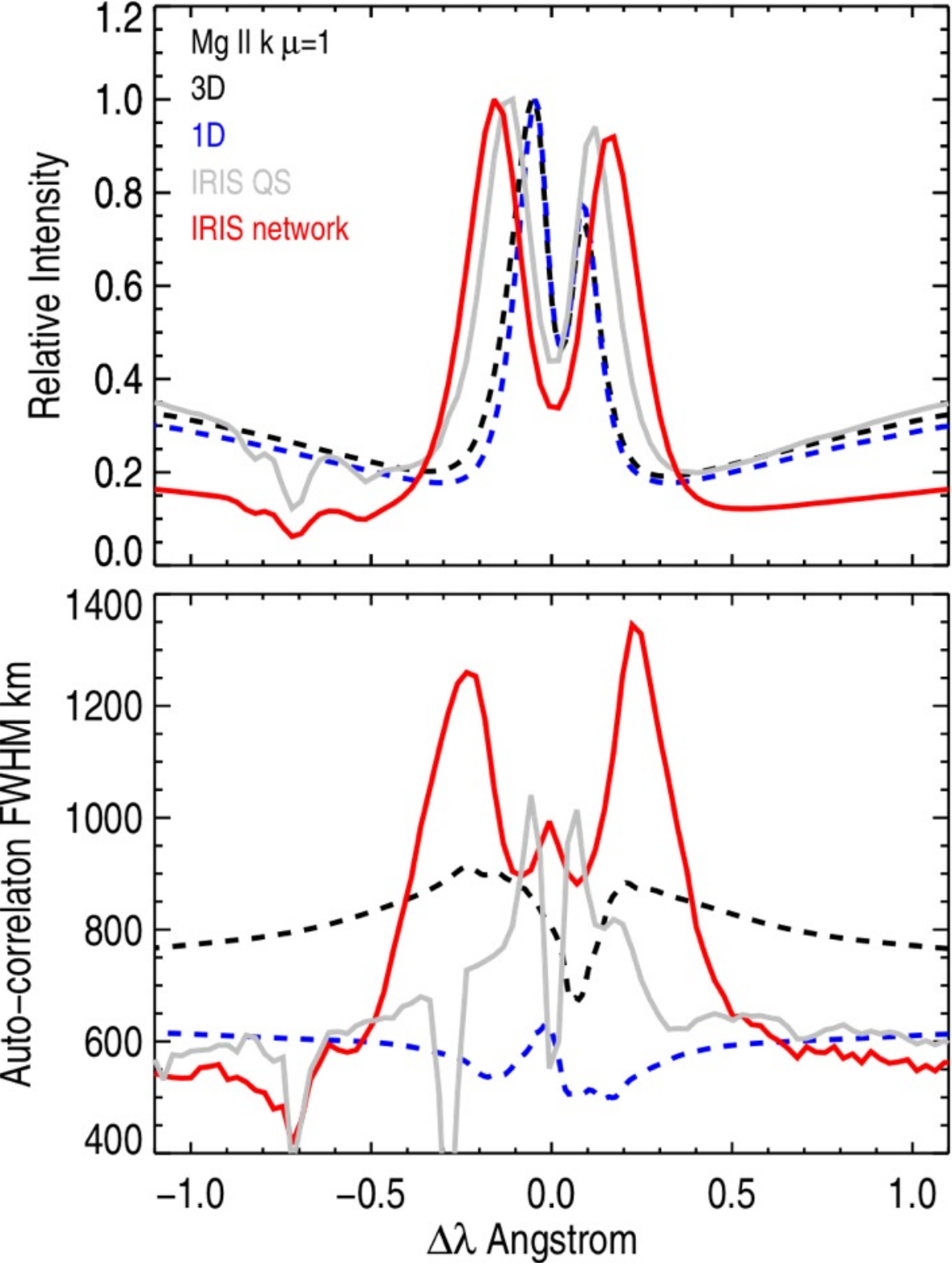}
  \caption{Average profiles (top panel) are shown with auto-correlation
    full-widths for the IRIS network and quiet regions (red and gray
    respectively) in the lower panel.
    The same data for the 3D and 1D calculations of
    \citet{2017A&A...597A..46S},
    are also shown as dashed lines.}
  \label{fig:autoc}
\end{figure}
}

\newcommand{\be}[1]{\begin{equation} \label{eq:#1}}
\newcommand{\ee}{\end{equation}}
\newcommand{\ba}[1]{\begin{eqnarray} \label{eq:#1}}
\newcommand{\ea}{\end{eqnarray}}
\newcommand{\kms}{km$\,$s$^{-1}$}

\newcommand{\solrad}{\ifmmode{R}_{\rm S}\else${R}_{\rm S}$\fi}
\newcommand{\solmas}{\ifmmode{M}_{\rm S}\else${M}_{\rm S}$\fi}

\newcommand{\tintu}{\ifmmode{\rm erg~cm^{-2}~s^{-1}sr^{-1}}\else 
  erg~cm$^{-2}$~s$^{-1}$~sr$^{-1}$\fi}
\newcommand{\fluxu}{\ifmmode{\rm erg~cm^{-2}~s^{-1}}\else 
  erg~cm$^{-2}$~s$^{-1}$\fi}
\newcommand{\velu}{$\,$km$\,$s$^{-1}$}

\newcommand{\wave}{\ifmmode{\lambda} \else$\lambda$\fi}

\newcommand\lta { \mathrel {\hbox to 0pt {\lower 3.7pt \hbox{$\sim$}
      \hss} \raise 1.7pt \hbox{$<$}}}
\newcommand\gta { \mathrel {\hbox to 0pt {\lower 3.7pt \hbox{$\sim$}
      \hss} \raise 1.7pt \hbox{$>$}}}

\newcommand{\philemail}{judge@ucar.edu}

\newcommand\new[1]{{\bf #1}}

\shortauthors{Philip Judge et al.}
\shorttitle{3D transfer}
\shorttitle{Line cores}

\slugcomment{}

\begin{document}

\title{Horizontal transfer versus spectroscopic turbulence}

\title{On the cores of resonance lines formed in the Sun's chromosphere}

\title{New light on an old problem of the cores of solar resonance lines}

\author{Philip G. Judge}
\affil{High Altitude Observatory,\\
       National Center for Atmospheric Research\footnote{The National %
       Center for Atmospheric Research is sponsored by the %
       National Science Foundation},\\
       P.O.~Box 3000, Boulder CO~80307-3000, USA; \philemail}

\author{Lucia Kleint}
\affil{Leibniz-Institut f\"ur  Sonnenphysik (KIS), Sch\"oneckstrasse 6, D-79104 Freiburg, Germany;\\
       University of Applied Sciences and Arts Northwestern Switzerland, Bahnhofstrasse 6, 5210 Windisch, Switzerland;
       0000-0002-7791-3241}


\author{Jorrit Leenaarts}
\affil{Institute for Solar Physics, Department of Astronomy, Stockholm University,
    AlbaNova University Centre, SE-106 91 Stockholm, Sweden}

\author{Andrii V.\ Sukhorukov}
\affil{Instituto de Astrof\'{\i}sica de Canarias,
    E-38205 La Laguna, Tenerife, Spain;\\
    Dpto.\ de Astrof\'{\i}sica, Universidad de La Laguna,
    E-38206 La Laguna, Tenerife, Spain;\\
    Main Astronomical Observatory, National Academy of Sciences of Ukraine,
    27~Akademika Zabolotnoho~St., 03143 Kyiv, Ukraine}

\author{Jean-Claude Vial}
\affil{Universit\'e Paris Sud, Institut d'Astrophysique Spatiale, UMR8617, 91405, Orsay, France;\\
       CNRS, Institut d'Astrophysique Spatiale, UMR8617, 91405, Orsay, France}

%
%
\def\red#1{{\color{red}{#1}}}

\begin{abstract}
We re-examine a 50+ year-old problem of deep central reversals predicted for
strong solar spectral lines, in contrast to the smaller reversals seen in
observations.
We examine data and calculations for the resonance lines of \ion{H}{1},
\ion{Mg}{2} and \ion{Ca}{2}, the self-reversed cores of which form in the upper
chromosphere.
Based on 3D simulations as well as data for the \ion{Mg}{2} lines from IRIS, we
argue that the resolution lies not in velocity fields on scales in either of the
micro- or macro-turbulent limits.
Macro-turbulence is ruled out using observations of optically thin lines formed
in the upper chromosphere, and by showing that it would need to have
unreasonably special properties to account for critical observations of the
\ion{Mg}{2} resonance lines from the IRIS mission.
The power in ``turbulence'' in the upper chromosphere may  therefore be substantially lower
than earlier analyses have inferred.
Instead, in 3D calculations 
horizontal radiative transfer produces smoother source functions, smoothing out intensity gradients in wavelength and in space.
These effects increase in
stronger lines.
Our work will have consequences for understanding the onset of the
transition region, the energy in motions available for heating the corona, and
for the interpretation of polarization data in terms of the Hanle effect applied
to resonance line profiles.
\end{abstract}

\
\keywords{Sun: atmosphere}

\section{Introduction}

During the 1960s and 1970s, with the advent of powerful numerical techniques for
performing non-LTE radiative transfer calculations, several groups vigorously
pursued modeling of critical spectral lines in an effort to understand the
structure and energy balance of the solar chromosphere.
Two striking problems quickly emerged.
The first concerned the wings of strong resonance lines\footnote{Here we define line ``cores'' and ``wings'' in terms of observed profiles, and
Figure~\ref{fig:problem} shows the \ion{Mg}{2} lines.
The core region lies roughly between the maxima in intensity marked historically
as ``$k_2$'' in the figure, the wings lie outside of these maxima.},
\new{which arose under}  the simplifying
approximation of complete redistribution (CRD): the wings were factors of
several too bright
\citep{1967AnAp...30..861D, 1973A&A....22...61L}.
The wing problem was resolved using the more realistic approach of partial
redistribution
\citep[PRD:][]{1973ApJ...185..709M,1974ApJ...192..769M,1976ApJ...205..874A}.
In this more realistic description, the emission and absorption processes in
spectral lines allow for frequency-by-frequency coupling between these processes
in the line wings, where scattering in the atomic reference frame is coherent.
The resolution of the wing discrepancy is illustrated by the qualitative
agreement between observations and the standard VAL3-C 1D model
\citep[][``VAL"]{Vernazza+Avrett+Loeser1981}
in Fig.~\ref{fig:problem}.
In this figure, the variations of the wing intensities with wavelength are
captured by these traditional models.
(The offset in intensity is within the calibration uncertainties of the rocket
spectra).

\figproblem

In the second problem, the centers of line cores ($k_3$ in the figure),
controlled by CRD, were computed to be too dim by factors of several, relative
to the $k_2$ peaks.
These discrepancies remain today for some of the strongest lines in solar and
stellar spectra, including H~L$\alpha$, and to a lesser extent L$\beta$
\citep{1978ApJ...225..655G,1979ApJ...230..924B};
the \ion{Mg}{2} $h$ and $k$ lines
\citep{Vernazza+Avrett+Loeser1981};
the \ion{Ca}{2} $H$ and $K$ lines
\citep{1981A&A...103..160L}.

Observed profiles of these strong lines were recognized to be asymmetric.
The explanation proposed early concerns the relative bulk motion between the
middle and upper chromosphere
\citep{1970SoPh...11..347A},
a result compatible with the upwardly-propagating radiating shocks calculated in
1D models by
\citet{Carlsson+Stein1995}.
This remains a related problem of interest here, but we do not address it
directly here.

More sophisticated calculations based upon 3D radiation-MHD (R-MHD) models
computed with the Bifrost code
\citep{2011A&A...531A.154G}
have reduced core $k_3/k_2$ \ion{Mg}{2} intensity ratio discrepancies somewhat
\citep{2017A&A...597A..46S},
and also for the $K_3/K_2$ \ion{Ca}{2} ratios
\citep{2018A&A...611A..62B}.
But issues in line cores \new{remained, in particular 
they found that observed peak separations implied that something was missing in their models.}

Physically, the cores of the strong lines generally correspond to wavelengths
where Doppler-shifted thermal and other motions control the opacity and source function.
Therefore in this ``Doppler core'', CRD is a reasonable approximation.
While the Doppler broadening widths are not known \textit{a priori}, the cores
are considered to span about 3~times the r.m.s. Doppler width, before
both opacity and source function become controlled by coherent
scattering.
In models and observations, this 3$\times$Doppler width varies systematically
from $\pm 25$~\velu{} for heavy elements such as calcium, to $\pm 50$~\velu{}
for hydrogen.

Given the importance of strong lines to the chromospheric energy balance, the
onset of the solar corona, the irradiance effects in the solar system and the
recent work on the Hanle effect in the cores of these lines
\citep[see, for example,][]{2018cosp...42E1564I},
we re-analyse various models and data-sets to understand better the meaning of
the remaining discrepancies between observation and models.

\figval

The early 1D models remain a reasonable first physical approximation because of
steep hydrostatic stratification, where the dominant direction of transfer of
radiative energy lies in the vertical direction
\citep[e.g.][]{2017ApJ...851....5J}.
However, even though in the 3D R-MHD models the stratification is steep (isobars
are more horizontal than vertical), there are many conditions where, owing to
magnetic and Reynolds stresses evolving in response to convection beneath, the
effects of horizontal radiative transfer are important. \new{Thus we study 
1D and 
3D~models, focussing on  the differences 
arising between 1D~formal versus 3D~formal solutions.}
Our attention is focused on the resonance lines of \ion{H}{1}, \ion{Mg}{2} and
\ion{Ca}{2}.

Lastly, a unique study by \citet{1994A&A...287..233B} reported simultaneous spectra
acquired with various $1\arcsec$-wide image-plane slits and different spectrograph
exit slits, in the resonance lines of \ion{H}{1}, \ion{Mg}{2} and \ion{Ca}{2},
from the LPSP instrument on the OSO-8 satellite.
They found that, while \ion{Mg}{2} and \ion{Ca}{2} data were similar, the
L$\alpha$ data appeared qualitatively different.

\section{Statement of the problem} 

Figures~\ref{fig:problem} and \ref{fig:val3c} illustrate our core problem of
interest: \textit{the calculated profiles have ratios between peak and core
brightness that systematically exceed observations, even those with spectral
resolutions of 200,000, by a factor of 1.5--2}.
Below we will examine in detail data of \ion{Mg}{2} from the Interface Region
Imaging Spectrograph \citep[IRIS,][]{2014SoPh..289.2733D} and from calculations
which highlight spatially-resolved spectra in contrast to the comparison of
spatial averages shown.

Armed only with 1-dimensional calculations, early resolutions that were proposed
to this problem were necessarily limited.
Within each calculation, the only option is to examine the assumed velocity
fields  within the chromosphere.
If included as a formal ``microturbulence'' (random fluid velocities exist on
scales below the photon mean free path), the effect is to increase the
wavelength differences between the  intensity peaks.
As shown in Fig.~\ref{fig:problem}, the separation-of-peaks for the $h$ and
$k$ lines in standard 1D calculations are close to spatially-averaged observed
values.
The well-known effect of variations in micro-turbulence is recalled in
Figure~\ref{fig:val3c}.
(Curiously, using the distribution of micro-turbulence of VAL3-C, the peak
intensities are split by similar amounts in wavelength, yet the line center
wavelengths are in the ratio $1 : 2.3 : 3.3$).

Another option in 1D modeling is to include line broadening as macro-turbulence,
then within each 1D calculation, the velocity would be introduced as a
macroscopic, not microscopic flow.
In this case, the emergent profiles, for changes small in magnitude compared
with micro-scale motions, introduce shifts (and asymmetries if velocity
gradients exist) of the entire core profiles.
This resolution, proposed by earlier authors
\citep[e.g.][]{1978ApJ...225..655G},
suggests that each observation sums spatially over intensities from several such
adjacent atmospheres in each pixel.
Figure~\ref{fig:cartoon} shows a schematic picture of this proposed solution.
Each atmosphere is a 1D solution to the non-LTE equations.
If each observation corresponds to a superposition of a set of 1D calculations
that are randomly distributed in line of sight velocity, to some degree the
differences can be reconciled.
Each observation then corresponds to the simple sum of individual theoretical
profiles, each shifted in wavelength according to a bulk velocity shift.

\cartoon

Spectroscopic turbulence was studied in the context of 1D~models by
\citet{1985cdm..proc..137C}.
They found that the regimes of applicability of the micro- and macro- turbulent
limits were found to be ${<}25$~km and ${>}3000$~km respectively, using velocity
fields with a variety of correlation lengths, for lines of \ion{Ca}{2}.
These lengths can be compared with a photon mean free path of 120~km and
thickness of the entire stratified chromosphere of 1500~km.
They showed that typical profiles could not be explained by a linear combination
of micro- and macro-turbulent flows.  

Three-dimensional radiation-MHD (R-MHD) models
\citep{2017A&A...597A..46S,2018A&A...611A..62B}
are based not on \textit{ad-hoc} explorations, but are solutions to the dynamic equations of motion and radiative transfer.
They are performed on grids fine enough to avoid too much dissipation, but for
which solutions can be obtained in reasonable computing time.
The calculations examined here have grids with spacings of 49 and 34~km
horizontally and vertically respectively, and are therefore capable of exploring
radiative transfer in different regimes of spectroscopic turbulence to the
extent that their intrinsic numerical smoothing permits significant changes in
velocities to occur between neighboring slices through the 3D~atmosphere which
determine the opacities and source functions.
These calculations were advanced in time using radiation transfer solutions based
upon the short-characteristics method, which is encumbered with a large,
unphysical diffusion.
\new{The source functions for strong lines are dominated by scattering.
As such, the modeled profiles, even computed with long  characteristics,}  must be treated with some caution.
Our final goal is to explore how profiles of strong lines
computed with the best 3D~MHD models compare with these observations.

\section{Observations and their analyses}

\subsection{Typical quiet Sun line profiles}

First, we examine the highest dispersion spectra of the three lines from the
literature, of the quiet Sun, but with low spatial resolution.
Figure~\ref{fig:problem} shows data of both the $h$ and $k$ lines from a
$7\arcsec\times130\arcsec$ area of the quiet Sun
\citep{1977STIN...7831029A}.
Figure~\ref{fig:val3c} shows profiles from regions of quiet Sun, highlighting
the cores of \ion{Ca}{2}, \ion{Mg}{2} and \ion{H}{1} resonance lines, from
several instruments.
The figure compares the observations with calculations from the 1D VAL-3C model.
Spectra of \ion{Ca}{2} lines were taken from the atlas of
\citet{1973apds.book.....D}.
The spectrometer used had a native spectral resolution of
${\cal R}= 600,000$, so that instrumental broadening is negligible.
The \ion{Mg}{2} spectra are taken from the ${\cal R}= 200,000$ quiet Sun spectra
shown in Fig.~\ref{fig:problem}
\citep{1977STIN...7831029A}.
Solar data for \ion{H}{1}~L$\alpha$ are all of a significantly lower spectral
resolution.
Those obtained from rocket flights or low Earth orbit are also affected by
absorption by neutral hydrogen in the Earth's upper atmosphere in the line core.
These include data from the HRTS-II experiment from
\citet{1991ApJS...75.1337B}
with ${\cal R}= 24,000$ and the LPSP instrument on OSO-8, which obtained quiet
Sun spectra with ${\cal R}= 60,000$.
Data from the SUMER instrument on SoHO have a lower spectral resolution of
15,000, even though the instrument, on the SoHO spacecraft at the L1~Lagrange
point, is unaffected by the geocorona.
L$\alpha$ SUMER data also have been obtained, mostly behind a mesh at the edge
of the detector, to avoid saturation.
\citet{2008A&A...492L...9C} obtained SUMER L$\alpha$ data behind the
partially-shut door of the telescope, of six quiet-Sun regions on
June 24--25, 2008.
Owing to the apodization of the pupil plane by the partly-closed door,
the angular resolution is difficult to assess, but they found the characteristic
asymmetry (blue peak brighter than red) which increased with brightness.
The central core was found to be typically 75\%{} of the brightness of
the blue peak, independent of limb distance, at a spectral resolution of 15,000.

Quiet Sun L$\alpha$ profiles from all three instruments are compiled in the
lower middle panel of Fig.~\ref{fig:val3c}.
The SUMER data shown there are from \citet{2001A&A...375..591C}.
All three lines show similar qualitative differences with 1D~calculations.
1D~calculations tend to predict brighter peaks and a darker central intensity.

A closer look at Figures~\ref{fig:val3c} and \ref{fig:cartoon}
immediately yields tight constraints on any macro-turbulence.
The right-most panels of Fig.~\ref{fig:val3c}, plotted against Doppler velocity,
show that the \ion{Ca}{2} $K$, \ion{Mg}{2} $k$ and \ion{H}{1} L$\alpha$ line cores 
would need systematically different macroturbulent velocity distributions to
explain the observed core-peak intensity ratios.
\textit{The \ion{Ca}{2} $K$ line would require r.m.s. speeds of
10~\velu, but \ion{Mg}{2} $k$ and \ion{H}{1} L$\alpha$ would need
r.m.s. values of 30 and 50~\velu{} for the superposition shown in
Fig.~\ref{fig:cartoon} to work.}
In stratified atmospheres, these three line cores form all within a region
termed the ``upper chromosphere'' (see, for example, Fig.~1 of
\citealp{2019A&A...623A..74D}
and Fig.~3 of
\citealp{2015ApJ...803...65S}).
We note that the area coverage of spicules inferred from data above the limb is
too small to contribute significantly to the spatially-averaged profiles,
\citep[e.g.][]{2010ApJ...719..469J,2016ARep...60..848M}.
However, models do show that the $k_2$ and $K_2$ peaks form substantially lower
in the atmosphere
\citep{Vernazza+Avrett+Loeser1981,2013ApJ...772...90L,2018A&A...611A..62B}.

The 3D~models have revealed that the peak separations of these lines can
increase in locations where the chromospheric temperature rise is located deep
in the atmosphere (see Fig.~20 of \citealp{2018A&A...611A..62B}).
But, no matter the details of the complex non-LTE formation of these
lines, we know of no first-principles reasons nor data that are compatible with
the idea of a systematic gradient of macro-scale motions that can rescue the
explanation that the peak-to-core ratios are small because of
macro-turbulence.
Indeed, below we show that the line profiles of \ion{Mg}{2} lines obtained with
the IRIS instrument, obtained with a high cadence and at the highest angular
resolution ever achieved, also can be used to reject this hypothesis.

\figsji

\subsection{Quiet Sun \ion{Mg}{2} line  profiles in time and space}

We examine detailed profiles for the \ion{Mg}{2} $h$ and $k$ lines from the IRIS
instrument
\citep{2014SoPh..289.2733D}.
While IRIS does not measure the UV~spectra at the highest spectral resolution,
the linear properties of the detector combined with the high angular resolution
make it uniquely suited to address the problem of interest.
Measurements in the \ion{Mg}{2} $h$ and $k$ lines have a spatial step along its
slit of $0.168\arcsec$, a critical sampling of the angular resolution of
$0.33\arcsec$.
Inspection of typical data (e.g. Figure~5 of \citealp{2014SoPh..289.2733D})
suggests that quiet regions have only  modest significant spatial
variations of the $k_2$ and $h_2$ peaks observed along the projected spectrograph
slit, as seen at the nominal angular resolution of IRIS.
The center-limb behavior shown in Fig.~6 of
\citet{2014SoPh..289.2733D},
reveals that peak separations first increase, and finally disappear some
$5\arcsec$ above the limb seen in the neighboring continuum.
The line profiles become a narrow single peak $5\arcsec$ higher.

All of the IRIS data were reduced to photometrically-calibrated spectra
using the IDL SolarSoft packages as well as deconvolved with the point spread
function (PSF) of the IRIS telescope, which has been measured during a Mercury
transit by \citet{2018SoPh..293..125C}.

Quiet Sun \ion{Mg}{2} data were obtained with IRIS close to disk center
beginning on February 27, 2014, at 5:39:28.850, in a ``sit and stare'' mode.
290 frames of 774 0.16635$\arcsec$-wide pixels (solar $y$) by
554 0.0254~\AA{}-wide wavelength pixels.
We will focus on spectra, but simultaneous IRIS slitjaw images were obtained in
the 1400~\AA{} channel.
Figure~\ref{fig:sji} shows examples of these images obtained 1000 and
3500 seconds into the time-series.
Each has been median-filtered over time (90~seconds) to remove internetwork
oscillations and fast dynamics in the low-density transition region.
\figyt
The thick lines show three regions we highlight below: network, haze, and
inter-network.
The color table saturates the highest DN to reveal UV~continuum emission that
forms in the lower chromosphere.

\figirismacrotk

Figure~\ref{fig:irisyt} shows wavelength-integrated intensities of the
\ion{Mg}{2}~$k$ line core, spanning 1~\AA{} (roughly the wavelength
separation between the $k_1$ minima), as a function of position along
the slit and time.
The data were acquired with a cadence of 16.8051 seconds, for a total duration
of 81 minutes.
The total solar area rotating under the fixed slit was therefore roughly
$13\arcsec\times 120\arcsec$ (solar $x$ and $y$ respectively).
This area covers 1~part in 2000 of the area of the solar disk, below
(Section~\ref{subsec:lucia}) we therefore examine a larger region.

\figiris

The data, when deconvolved, have a sufficiently high angular resolution
($0\farcs33$) that the macro-turbulent picture can be directly addressed.
Each resolution element (two pixels) corresponds to about 240~km on the solar
surface.
If a superposition such as that shown in Fig.~\ref{fig:cartoon} were responsible
for filling-in the line cores, then some profiles adjacent in space should
exhibit large variations between separate resolution elements, i.e. on scales
down to 240~km.
But such variations are absent, as shown in the typical, randomly chosen samples
of line profiles of network, inter-network, and a hazy intermediate region.
These profiles are typical of the entire data-set, including the $h$-line.

\subsection{Spatial auto-correlations}

The leftmost panel 
of 
Figure~\ref{fig:irismacrotk} 
has not been spatially de-convolved, the others have.
The tick-marked line in the rightmost panel shows ten spatial pixels, \new{each of} which
critically sample the convolved data.
Very little structure below scales of about 5 pixels ($\equiv 600$ km) is visible.
This figure prompted us to measure quantitatively the spatial structure along
the slit of the entire dataset, using auto-correlations.
Thus, we computed the spatial auto-correlation lengths along the IRIS slit at
each wavelength, and time, for those columns in Fig.~\ref{fig:irisyt}
corresponding to all three marked regions.

The figures below show autocorrelation lengths of the intensity images
$I_\lambda(y)$, which are the characteristic full-widths at half-maximum (FWHM)
of the functions
\begin{equation}
    c_\lambda(\ell) = 
         I_\lambda(y) \cdot I_\lambda(y+\ell),
\end{equation}
centered on the maximum ($\ell=0$).
Since $c_\lambda(\ell)$ depends on each wavelength $\lambda$ across the line
profiles, any variations of line-of-sight velocity on resolvable scales
will be reflected by changes in the monochromatic intensities on the same length
scales, thereby imprinting the velocities onto these correlation lengths.
Figure~\ref{fig:iris} shows these $c_\lambda(\ell)$ values.
Over-plotted is the average line profile in blue, and the autocorrelation FWHM
lengths averaged over space and time, in white, along the network, haze, and
inter-network slit regions.

Small values of the FWHM of $c_\lambda(\ell)$ correspond to rapid changes in
space, and \textit{vice versa}.
An obvious feature of Fig.~\ref{fig:iris} are that in the wings of the
$k$ line ($\lambda$ outside of the $k_1$ minima) the three regions are not
easily distinguishable, unlike the line cores which are radically different.
There are some extended periods where the structure in the wings of the network
contain consistently small FWHM values (between 1400 and 1600, or 1800 to 2200
seconds for example), these upper photospheric signatures behave, statistically
at least, in a similar fashion.
The average autocorrelation lengths outside of line cores are all close to
400~km (the averages are shown as white lines in the figure).

But within the cores, the network and other regions behave quite differently.
The average intensity profiles in the haze and internetwork regions are similar
in shape to the average correlation length profiles (white and blue lines).
But in the network, the very cores of the autocorrelation profiles possess a
peak at the $k_3$ core that is absent in the intensity profiles.
The spatial autocorrelations are also larger farther from line center
($\Delta\lambda > 0.1$~\AA{} in the network region, with the exception of a short
period around $t=2600$ seconds (left panel of Fig.~\ref{fig:iris}).
During the first half of the time series autocorrelation lengths can exceed
${\approx}2500$~km near the $k_2$ peaks, and almost symmetric about line center.
Later, after about 3000 seconds the region of network shows an asymmetric
autocorrelation profile in wavelength, line cores lengths exceeding 2000~km, but
with a coherent darker feature, i.e. smaller lengths between 0.1--0.2~\AA{} to
the red of line center (Doppler redshifts of 10 and 20~\velu) between 3200 and
4300 seconds.

All three regions have average spatial scales exceeding 800~km at all
wavelengths which have significant core emission (within about ${\pm}0.3$, 0.2,
and 0.1~\AA{} respectively for the network, haze, and internetwork).
The spatial auto-correlations within the cores (${\pm}0.5$~\AA) have structures
with a median of 1060~km, mean of 1140~km, and r.m.s. variation of
440~km, for the haze and internetwork data, with 20\%\ higher typical values for
the network.

In contrast, the haze and internetwork regions possess a significant and
persistent dark streak within 5--8~\velu{} (Doppler shift) of $k_3$.
There, the spatial FWHM correlation lengths fall to 400~km, which is only about
twice the spatial resolution of IRIS.
We will speculate on the origin of these small structures below.

In summary, these IRIS data reveal that:
\begin{enumerate}
\item Away from the line cores (outside $k_1$ minima), all regions have
    autocorrelation FWHM values close to 400~km, roughly twice the IRIS
    resolution.
\item At and near wavelengths of intensity maxima ($k_2$), FWHM 
measurements 
exceed
    1000~km, increasing to over 1500~km in network regions.
\item Outside of network regions, $k_3$ has a persistent small spatial FWHM of
    $\approx 400$~km, seen only in line center pixels and those two immediately
    to the red side (${\le}10$~\velu{} from line center).
\item Over network regions, autocorrelation lengths are systematically larger,
    extend further in $\Delta\lambda$, and have a broad minimum near 800--900~km
    across the central ${\pm}0.2$~\AA.
    Within ${\pm}0.1$~\AA{} the lengths increase producing a third, central small
    peak in average autocorrelation profiles.
\end{enumerate}
To explain the small $k_3/k_2$ ratios, the macro-turbulent picture requires
large changes in line-of-sight velocities on scales down to the 200~km
resolution limit of IRIS.
Our analysis of typical IRIS data for \ion{Mg}{2} rejects this hypothesis.

\subsection{Other relevant observations}

Other spectral lines have been observed at a high angular resolution
which can also shed light on conditions where the cores of strong solar
resonance lines form.
Spin-forbidden lines of \ion{O}{1} at 1356 and 1358~\AA{} have been observed
repeatedly with UV spectrometers.
The lines are optically thin across most of the chromosphere, the shared upper
level is 9.15~eV above the ground level of \ion{O}{1}.
The \ion{O}{1} ionization is strongly tied to the \ion{H}{1} level populations
through a well-known resonant charge-transfer reaction, and so excitation by
collisions with electrons will occur under conditions common to these lines of
\ion{O}{1} and L$\alpha$, whose upper level lies at 10.19~eV.
Naturally, the L$\alpha$ line source function and emergent intensity are
dominated by photon scattering, unlike the spin-forbidden lines of \ion{O}{1}.
But the two lines otherwise are excited under similar conditions within the
 chromosphere.

The HRTS instrument has provided spectra near 1350~\AA{} \new{at a spectral resolution 
(50~m\AA) similar to IRIS, but a lower spatial (1$\arcsec$) resolution.}
HRTS data analyzed by
\citet{1989ApJ...346..514A}
show that this line's kinematics is unspectacular, with r.m.s. Doppler
shifts of 2~\velu, and linewidths close to that of the instrument.
The unresolved motions are at most 40\%{} of the sound speed, which is
10--15~\velu.
It seems there is little power in motions, resolved or unresolved, across the
quiet Sun's upper chromosphere.
\citet{2015ApJ...809L..30C}
reported widths of ${\sim}7$~\kms of the \ion{O}{1} 1356~\AA{} line in IRIS data
in quiet and plage regions. 
\new{Optically thin lines such as the 1356 line have contributions from many scale heights across the chromosphere, so their line widths reflect
the line-of-sight sum of many different heights.}

\subsection{A broader sample of \ion{Mg}{2} IRIS data}
\label{subsec:lucia}

The above IRIS dataset was selected to represent a typical quiet Sun
region at disk center.
But these data sampled just 1 part in 2000 of the solar surface.
We therefore looked at statistical properties of the IRIS \ion{Mg}{2} $h$ and
$k$ profiles from a spatial (raster) scan, that of April 15, 2014,
between 05:25:24 and 05:43:28 UT.
This spanned an area of $127\arcsec \times 119\arcsec$, which is ten
times larger than the sit-and-stare observation.

This observation included a little more active Sun, lying at the periphery of a
substantial active region (NOAO 12036).
We found that only about 10\%{} of the line profiles had $k_2/k_3$ intensity
ratios between 4 and 6, spanning the value of ${\approx}5$ for the VAL3-C
calculation.
The bright chromospheric network showed no more spatial correlation with these
deep line ratios than other pixels in the instrument's field of view.
On balance, these data from a larger area confirm that large $k_2/k_3$ intensity
ratios above 4 are present at a 10\%{} level.
The median ratio is 2.5. 

\section{Re-analysis of recent 3D calculations}

We have examined model data from \citet{2017A&A...597A..46S} and
\citet{2018A&A...611A..62B}.
The atmospheric structure in the models is the same in both articles and is
determined by  R-MHD calculations, using the short
characteristics method to solve for radiation gains and losses in the energy
equation.
After the calculations have been evolved, the authors then studied  solutions
to the transfer equation using long characteristics for the 1D and 3D
calculations.
Here we focus on the computations for the $k$ line by
\citet{2017A&A...597A..46S},
and also comment on those for the \ion{Ca}{2} $K$ line at 393.4nm.
Our purpose here is to examine the hypothesis that the line cores of
chromospheric resonance lines are filled in by horizontal radiative transfer.

The horizontal component of the numerical grid has pixels of width 47~km.
While the MHD calculations upon which these radiative transfer calculations are
based are themselves highly diffusive compared with the real Sun, the thermal
structure in their calculations nevertheless has features down to these scales.
The computational data analyzed here have been re-binned to scales half of the native
resolution, 98~km in the horizontal direction.
This scale is comparable or smaller than typical photon mean free paths through
the stratified chromosphere.
While a finer grid is desirable, these calculations are fine enough to reveal
the effects explored below.

The modeled region is similar to  enhanced network on the Sun, with
an unsigned magnetic field strength of 50~G passing through the photosphere
\citep{2016A&A...585A...4C,2017A&A...597A..46S,2018A&A...611A..62B}.
A typical snapshot of a time-dependent 3D calculation is taken.
From the source functions in each voxel, the long-characteristic method is used
to produce the emergent intensities, including PRD.
Therefore the solutions shown are a kind of hybrid: the source functions and
the optical depth scales are computed using the diffusive
short-characteristic method, but the final emergent intensities, given these
parameters, are far less diffusive.

Figure~\ref{fig:mgii} shows emergent intensities from a small part
($6\times6$ Mm$^2$) of the calculated area ($24\times24$ Mm$^2$), using the long
characteristic method, for a region at disk center ($\mu=1$).
When compared with observations, we note firstly, that the computed features
$k_{2V}$ and $k_{2R}$ are factors of between one and two brighter than the
average quiet Sun data shown.
This is to be expected given the higher concentration of the magnetic
field used in the computations.
But the $k_2/k_3$ contrasts are also significantly higher, both in 3D and 1D,
and the $k_{2V}$ and $k_{2R}$ separations are smaller than observed.
Qualitatively similar results (from \citealp{2018A&A...611A..62B} but not shown
here) are seen for the \ion{Ca}{2} $K$ line, where the average $K_{2R}$
and $K_{2V}$ peaks are separated by 0.33~\AA{} in both quiet and plage regions
\citep[compare Figures~3 and 5 of][]{Linsky+Avrett1970},
whereas the average over the computed enhanced network region is about half of
this (0.17~\AA).

\figmgii

Significantly, in comparison with plage observations, these $k_3$ and $K_3$
components are far deeper in the computations.
This suggests that the source function is on average larger where this feature
is formed than is captured in the calculations.

In Figure~\ref{fig:autoc} we show spatial auto-correlation data computed from
the 3D and 1D intensities of the $k$ line, together with spatially-averaged
intensity profiles.
Solid lines show observations, dashed lines computations.
The auto-correlation characteristic lengths show three kinds of behavior
depending on the wavelengths relative to line center.
In the line wings, the images have the smallest spatial scales, with FWHM values
of about 750 and 610~km.
In 3D, these scales steadily increase towards the line cores, peaking near the
observed peaks $k_{2V}$ and $k_{2R}$.
(In 1D these lengths decrease towards the line center before showing a small
increase at $k_{2V}$ wavelengths).
Within the line core, the widths computed in 3D drop dramatically to below
700~km.

\figautoc

The comparison of 3D and 1D images in Fig.~\ref{fig:mgii} is instructive in ways
complementary to the interesting points made by
\citet{2018A&A...611A..62B}.
Figure~\ref{fig:autoc} shows clearly that 3D~transfer effects are
important within $\pm 1$~\AA{} of line center: the FWHM length scales are larger
on average by a factor of two for the $k$ line (15--20\%\ for the $K$ line) in
3D than 1D.
The 3D features all appear ``fuzzier'' to the eye than their 1D counterparts, on
scales of a few hundred kilometers in Fig.~\ref{fig:mgii}.
We will speculate further on what is missing in these calculations in
Section~\ref{sec:discussion}.

The second and fourth rows of Fig.~\ref{fig:mgii} show how the calculated model
might be observed through filter instruments, with FWHM values of 0.1, 1, and
3~\AA.
Wider FWHM instruments have been used in solar physics for many years to record
the \ion{Ca}{2} lines, and other, narrower filter widths have been developed%
\footnote{see, e.g. 
\url{https://www.su.se/isf/}}.
These panels suggest that 3D smearing effects are biggest at wavelengths within
the emission cores.
The differences are probably lower limits, given that the source functions for
strongly scattering lines are smeared by the use of the short characteristics,
which will tend to underestimate differences between the 1D and 3D formal
solutions to the transfer equation.

\section{Discussion}
\label{sec:discussion}

We have re-visited an old problem concerning the upper chromosphere of the Sun:
is there a clear discrepancy between observed and computed profiles in the
Doppler cores of the strongest lines?
If so, can we determine its origin?
Firstly, we compiled data to re-examine literature from the 1970s when this
problem was first addressed.
Using these data, including optically thin lines generated in the same regions
as the strong lines, we confirmed earlier work showing that micro-turbulence
cannot account for the systematic behavior of the data and argued that
macro-turbulence also fails.
On this basis, we then examined the $k$ line of \ion{Mg}{2} in detail, taking
advantage of the quality and angular resolution of the IRIS instrument.
We then explored radiation MHD simulations of an enhanced network
region of the $K$ line of \ion{Ca}{2}.
In particular, we compared the line cores in 1D and 3D to explore the effects of
 horizontal radiative transfer on the emergent spectra generated
throughout the chromosphere.

Our first conclusion is that \textit{the small peak-core ratios in these
resonance lines cannot be due to turbulence on any scale}.
Given the well-known result that micro-turbulence serves mostly to increase the
separation of peaks (see Fig.~\ref{fig:val3c}), our conclusion rests on the
refutation of macro-turbulence.
Three lines of argument all point to this result.

\begin{itemize}
\item The macro-turbulence limit requires the spatial averaging of profiles of
    fluid elements, which are Doppler-shifted by a velocity distribution,
    which spans the computed 1D separation of peaks.
    If the distribution were narrower, little difference would be seen between
    the 1D and average profiles.
    If larger, then we would see multiple peaks for a small number of averaged
    elements, and/or a severely broadened feature.
    But observations clearly show a steady increase of separation of peaks from
    \ion{Ca}{2} to \ion{H}{1} by a factor of three in Doppler shift.
    Yet the \new{deep} cores of these features all form within the confines of the upper
    chromosphere and higher. \new{Whether or not the chromosphere is thermally isolated from the corona above by mostly horizontal magnetic fields, these line cores  likely form in regions of very steep gradients in electron  temperature where ionization causes them to become optically thin.}

\item Optically thin UV lines observed with the HRTS instrument since the 1970s
    also \new{have contribnutions from} the upper chromosphere.
    Spin-forbidden lines of \ion{O}{1} near 1356~\AA{} require almost the same
    population of electron energies to be excited as H~L$\alpha$, yet they are
    spectrally unresolved by HRTS and show small subsonic deviations of just a
    few \velu{} along the instrument's slit
    \citep[e.g.,][]{1989ApJ...346..514A}.
\item Spatially de-convolved profiles of the \ion{Mg}{2} resonance lines
    observed by IRIS show very few features smaller than 600~km along the slit,
    yet the spatial resolution is close to 240~km (Figures~\ref{fig:irismacrotk}
    and \ref{fig:iris}).
    Between the $k_1$ minima, average autocorrelation lengths vary between
    800--1200~km, larger values being seen in brighter regions of the network.
\end{itemize}

For our second conclusion, based upon differences between the 1D and 3D
autocorrelation calculations (Figure~\ref{fig:mgii}), \textit{at least some of
the peak-to-core ratio discrepancy is due to horizontal components of
radiation transfer}.
Figures~\ref{fig:mgii} and \ref{fig:autoc} unambiguously reveal the small-scale
smearing effects of horizontal radiative transfer in the emergent intensities.
Horizontal radiative transport in space is the \new{most likely way to explain the difference} between the
1D and 3D calculations.
These differences cause the lowering of contrast across the line profiles in
both space and frequency, and larger the autocorrelation lengths in space.



We can also draw a third conclusion, namely that the upper chromosphere may well
harbor \textit{less ``turbulence'' than previously thought}.
The profiles of the three strongest lines formed in this region simply do not
support supersonic motions within the uppermost layers of the stratified
atmosphere for the vast majority of the time.
In this picture, supersonic spicules and other features simply are not abundant
enough and do not cover enough area of the surface to be significant, on
average, in affecting these line profiles.

\subsection{Further speculations}

Beyond these conclusions, the confrontation between computed and observed
profiles necessarily becomes more speculative.
Arguably, modern MHD calculations capture the lower chromosphere better than the
upper chromosphere and higher layers.
After all, the energy in higher layers is modified and filtered by its
propagation and dissipation through the lower layers, which themselves remain a
subject of active research.
The interface between the cool chromospheric and hot coronal plasma is not only
difficult to model accurately, but observations over the past century have
consistently revealed the complex thermal nature of the fine structure
of the upper chromosphere.
Even the latest radiation MHD calculations attempting to understand why the Sun
must produce spicules are based not on 3D but ``2.5D'' calculations in which
special symmetries are imposed \citep{2018ApJ...860..116M}.
It remains premature to state that we really know what spicules, and other
products of the plasma dynamics of the chromosphere extending higher up, really
are.

These and other structures entirely absent in 1D~calculations and
3D~calculations may well influence in our analysis.
As made explicit by
\citet{2015SoPh..290..979J},
such structures can have optical depths and source functions \textit{radiatively
de-coupled} from the chromosphere beneath.
These structures will simply absorb or emit radiation along the line of sight on
their physical scales, essentially disconnected from the non-locally coupled
radiative transfer solutions leading to the bulk of the line profiles beneath.
They can have structure on scales much smaller than the photon path
lengths.
Owing to the magnetic control of these plasmas owing to lower densities and
pressures, they might appear at almost any Doppler shift up to the (high)
Alfv\'en speed.
Thus we might expect to see absorbing and/or emitting structures perhaps with
length scales down to the diffraction limit if such structures cover much of the
chromosphere much of the time.
This kind of picture might explain the peculiar structures seen in the network
panel in the upper part of Fig.~\ref{fig:iris}.
In the internetwork and haze regions, we see the smallest structures in the line
cores, mostly within $\pm 0.05~\text{\AA} \equiv \pm 5$~\velu{} Doppler shifts
(Figure~\ref{fig:iris}, seen in the images and the column-averaged white lines).
These are sub-sonic speeds.
Super-sonic motions associated with spicules and some fibrils are notable by
their absence.
For a typical length of a spicule of 5--10$\arcsec$, and assuming that $10^5$
such spicules are present at any time on the Sun
\citep{2016ARep...60..848M},
we would expect to see one spicule cross a randomly oriented slit of this length
at any time.
The small-scale features at line-center observed almost everywhere are therefore
not related to spicules.
Interestingly, similar features are present in the numerical models for the
\ion{Ca}{2} $K$ line, outside of the chromospheric network.

Finally, as noted earlier, if the location of the chromospheric temperature rise
is deeper in the atmosphere than typical models suggest (see Fig.~20 of
\citet{2018A&A...611A..62B}),
then this problem should be re-visited.
Hints of a potential contribution of such a regime to a new understanding are
present in line widths computed from A--F models in
\citet{Vernazza+Avrett+Loeser1981},
and the calculations of
\citet{2015ApJ...809L..30C}.
This is beyond the scope of the present paper but will be addressed in future.

\section{Conclusions}

Our  results listed above  suggest that the amount of ``turbulence'' present in the
upper chromosphere has been over-estimated in prior work.
This may have significant consequences for the energy budget that can be used to
heat the overlying regions of the solar atmosphere.

Our analysis is incomplete in many ways: for example, we have no pure solar
spectra for H~L$\alpha$ with sufficient spectral resolution and photometric
quality to make comparisons with computations; neither have we any images with
sufficiently narrow passbands (Doppler widths ${\lta}10$~\velu{}) to be able to
compare with predictions from computations including scattering such as
\andrii{in} Fig.~\ref{fig:mgii}.
Even so, it is clear that the various models in 3D as well as 1D are missing
essential physical processes affecting the typical conditions in the upper solar
chromosphere.

The consequences of revisiting of early work remain to be fully understood.
There are many observations relevant to this study, such as the remarkable
images of broad-band L$\alpha$ light from the VAULT instrument
\citep[e.g.][]{Patsourakos+Gouttebroze+Vourlidas2007},
of chromospheric fine structure in very \andrii{narrowband} images in H$\alpha$,
the \ion{Ca}{2} infrared triplet lines and others
\citep[e.g.][]{2014ApJ...785..109L}.
Our work is almost certainly important in terms of the amount of energy stored
in and transported by the small-scale mass motions that we need to invoke, as
well as 3D transfer effects, to help explain the problem we have addressed.

\acknowledgements

J.L.\ acknowledges support through the CHROMATIC project (2016.0019) funded by
the Knut and Alice Wallenberg foundation.
The Institute for Solar Physics is supported by a grant for research
infrastructures of national importance from the Swedish Research Council
(registration number 2017-00625).
A.V.S.\ acknowledges financial support from the European Research Council (ERC)
under the European Union's Horizon 2020 research and innovation programme (ERC
Advanced Grant agreement No~742265) and from the Swiss National Science
Foundation (SNSF) through Grant CRSII5\_180238.

NCAR is sponsored by the National Science Foundation.
\emph{IRIS} is a NASA small explorer mission developed and operated by LMSAL
with mission operations executed at NASA Ames Research center and major
contributions to downlink communications funded by ESA and the Norwegian Space
Centre.
Computations presented in this paper were performed on resources provided by the
Swedish National Infrastructure for Computing (SNIC) at the National
Supercomputer Centre (NSC) at Link\"oping University, at the PDC Centre for High
Performance Computing at the Royal Institute of Technology in Stockholm, and at
the High Performance Computing Center North (HPC2N) at Ume{\aa} University.

\bibliographystyle{apj}
\bibliography{ms.bib}

\end{document}